\documentclass{article}

\usepackage{microtype}
\usepackage{graphicx}
\usepackage{subfigure}
\usepackage{booktabs}
\usepackage{amsmath} 
\usepackage{amssymb}
\usepackage{hyperref}

\usepackage[accepted]{icml2019}

\begin{document}

\twocolumn[

\title{Searching, fast and slow, through product catalogs}
\date{\vspace{-0.2in}}
\maketitle

\icmlsetsymbol{equal}{*}

%\begin{icmlauthorlist}
%\icmlauthor{Laurent Bou\'e}{MS}
%\end{icmlauthorlist}

\begin{icmlauthorlist}
\icmlauthor{Dayananda Ubrangala}{MS}
\icmlauthor{Juhi Sharma}{MS}
\icmlauthor{Sharath Kumar Rangappa}{MS}
\icmlauthor{Kiran R}{MS}
\icmlauthor{Ravi Prasad Kondapalli}{MS}
\icmlauthor{Laurent Bou\'e}{MS}
\end{icmlauthorlist}

\icmlaffiliation{MS}{Microsoft, CX Data Cloud + AI}

\vspace{0.4in}

\begin{abstract}

\vspace{0.4in}

String matching algorithms in the presence of abbreviations, such as in Stock Keeping Unit (SKU) product catalogs, remains a relatively unexplored topic.  In this paper, we present a unified architecture for SKU search that provides both a real-time suggestion system (based on a Trie data structure) as well as a lower latency search system (making use of  character level TF-IDF in combination with language model vector embeddings) where users initiate the search process explicitly.  We carry out ablation studies that justify designing a complex search system composed of multiple components to address the delicate trade-off between speed and accuracy.  Using SKU search in the Dynamics CRM as an example, we show how our system vastly outperforms, in all aspects, the results provided by the default search engine.  Finally, we show how SKU descriptions may be enhanced via generative text models (using gpt-3.5-turbo) so that the consumers of the search results may get more context and a generally better experience when presented with the results of their SKU search.

\vspace{0.4cm}

\end{abstract}
\vspace{0.4in}
]

\icmlcorrespondingauthor{\\ Laurent Bou\'e}{laboue@microsoft.com}
\printAffiliationsAndNotice{}

\section{Introduction}

Having a dependable and efficient search engine embedded within a Customer Relationship Management (CRM) system is vital. It empowers sales teams and support staff to swiftly retrieve pricing, availability, and other product details, enabling them to respond accurately and promptly to customer queries. Unfortunately, due to their multifaceted nature, the search functionality of CRMs often leans towards simplicity.  The resulting reduced search accuracy is especially pronounced when users seek information with complex abbreviations as is a common scenario for searching through Stock Keeping Units (SKUs) catalogs.

In this work, we use the Dynamics CRM as an example of where SKU search is performed daily by thousands of Microsoft sellers.  At the time of writing, the product catalog contains~$\approx~87,000$ active SKUs from multiple pricelists.  Our goal is to describe the multiple components we have put into place to design a comprehensive and production-ready search system that not only supports abbreviated strings but also offers a vastly superior performance and user experience to that of the existing Dynamics CRM.

We start in Section~\ref{sec:data} by describing a minimal dataset of features one may use for SKU search.  Next, we realize that a useful search engine requires the capability to handle both dynamic search (which provides instant results as the user types each character) and complete search (activated when the user hits the search button).  Dynamic search offers rapid feedback allowing users to refine their queries on the fly.  In Section~\ref{sec:Suggestions}, we show how dynamic search may be solved using a Trie data structure.  We continue in Section~\ref{sec:completeSearch} by showing how top accuracy may be achieved by combining character-level TF-IDF with language model embeddings together.  This leads to Section~\ref{sec:Architecure} where we describe the overall search system that strikes a balance between speed and accuracy.  Finally, we show in Section~\ref{sec:GPTgen} how the SKUs descriptions' may be enhanced using a GPT-based generative text model.

\section{Related work}
\label{sec:relatedWork}

Currently the search functionality of the Dynamics CRM is provided by invoking custom SQL queries that almost entirely rely on the \texttt{contains} keyword~\cite{SQLcontains}.  Although this predicate may be extended to support some form of fuzzy matching, in practice, the results have not been satisfactory with many queries.

The Dynamics CRM search receives an average of $\approx 11,000$ daily search requests.  Among those, about~$\approx 18\%$ return zero results while another~$\approx 35\%$ return an unreasonably high number of~50 or more product suggestions.  Currently, these search results are not sorted based on any criteria. Instead, they consist of randomly selected results returned by the \texttt{contains} predicate.  Furthermore, the average response time is rather slow at~$\approx 1$ second.

We start in Section~\ref{sec:data} by describing the set of features we extracted from the Dynamics CRM.  Those features are kept to a minimum and not specific to Dynamics so that other CRMs should also benefit from this paper.  Then we move on to the 3 main modules that define our search system: part number pattern matching (Section~\ref{sec:patternMatch}), dynamic suggestions (Section~\ref{sec:Suggestions}) and user-initiated search (Section~\ref{sec:completeSearch}).  Next, we present the fast and slow trade-off between accuracy and latency.  Finally, we conclude in Section~\ref{sec:GPTgen} with a description of a generative text module based on ChatGPT text completion to generate helpful SKU description.

\section{Stock Keeping Unit dataset}
\label{sec:data}

\subsection{Minimal set of product features}

\begin{figure*}[ht]
\begin{center}
\centerline{\includegraphics[width=2\columnwidth]{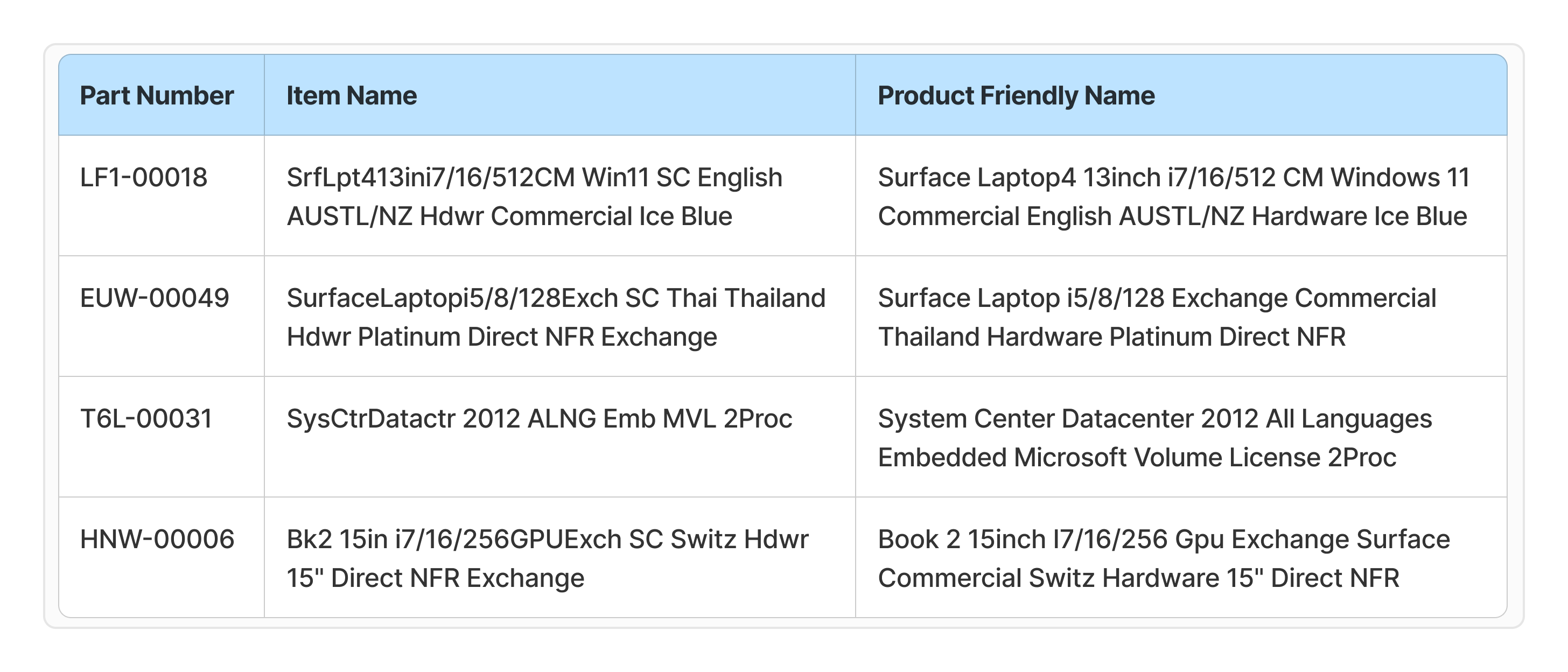}}
\caption{Example samples of the minimal set of features we are considering for the SKU product search system.  More details about the part number matcher can be found in Section~\ref{sec:patternMatch} and about how product friendly names are generated in Section~\ref{sec:abbreviationExpansion}.}
\label{fig:features}
\end{center}
%\vskip -0.4in
\end{figure*}

The dataset can be presented in a tabular format of approximately~$87,000$ rows where each row represents a product present in the catalog.  The minimal set of features characterizing a Stock Keeping Unit (SKU) is composed of:
\begin{itemize}
    \item Part number.  Those are explained in~Section~\ref{sec:patternMatch}.
    \item Item name. This is a highly abbreviated name which is hard to understand for any new reader. This has been like this for several years because of constraints in the legacy systems, for example on the length of the Item Name in the UI and the length constraints in the backend database systems.
    
    \item Product friendly name.  This is an expanded name for the SKU for easy understanding and will greatly help sellers in picking the right SKUs. However, these expanded names are available for approximately~30\% of the products and our abbreviation expansion strategy is described in~Section~\ref{sec:abbreviationExpansion}.
\end{itemize}

Those~3 fields are typically the ones that sellers would be searching for in their CRM queries and some examples are displayed in Fig.~\ref{fig:features}.  Obviously other SKU attributes (such as product family name, release dates and others...) may be added to have an even more accurate description of the SKUs.  For the purpose of this paper, we focus only on those fields that are most likely to exist in the catalog of any SKUs in any CRMs.

\subsection{Expanding abbreviations}
\label{sec:abbreviationExpansion}

As mentioned in the previous section, item names often contain rather cryptic and domain-specific abbreviations.  In many cases, sellers are not even aware of the meaning of those abbreviations and therefore are unlikely to use them in their search queries.  The process of expanding those abbreviations to turn normal item names into their ``friendly'' names requires extensive exploratory data analysis (EDA) of the catalog.  

First of all, with the help of domain experts, we have compiled a dictionary of~$\approx 540$ known abbreviations and their expansions.  However, the main difficulty lies in the way that item names are defined without separator characters between the abbreviations.  As an example, a typical item name shown in the first row of~Fig.\ref{fig:features} may start with
\begin{equation*}
\texttt{SrfLpt413ini7/16/512}
\end{equation*}
Given this string, we have devised a custom algorithm that splits the above example into a series of fields \{\texttt{SrfLpt4} , \texttt{13in} , \texttt{i7}, \texttt{16}, \texttt{512}\}.  Essentially, the algorithm heavily relies on the catalog statistics revealed by our extensive EDA to which we apply very specific string matching splits.  This algorithm is successful at parsing about~10\% of the catalog. 

The next step in our abbreviation expansion strategy consists in going over all the item names and the fields extracted from the first step (when applicable) to look for strings that follow a camel case convention~\cite{camelCase}.  Those are good candidates for further splitting into abbreviated terms.  For example, \texttt{SrfLpt4} may be split into \{\texttt{Srf}, \texttt{Lpt}, \texttt{4}\} under this captital letter assumption.

Finally, we look up those possible abbreviated terms in our known dictionary of abbreviations to look for exact matches. Continuing on the above example, we would then expand \texttt{Srf} into \texttt{Surface} and \texttt{Lpt} into \texttt{Laptop}.  This expansion can be seen in the last column of~Fig.\ref{fig:features}.  Overall, this approach allows us to expand about~30\% of the item names into so-called product friendly names.

We reserve a more systematic treatment of abbreviation expansion and discovery to a future extension of the model~\cite{autoAbbreviatons1, autoAbbreviatons2}.

\section{Part number pattern matching}
\label{sec:patternMatch}

One common search scenario is for sellers to copy / paste what they think is a good part number directly into the search box.  So what we implemented first is a pattern matching system where we classify the search queries to see if they potentially match a part number template.  If they do, then even if the part number that the seller is searching for is no longer active, there may be other part numbers that are closely related and we show these ones as the search results.

Typically part numbers follow established conventions depending on the specific CRM we are considering.  In the case of the Dynamics CRM, part numbers are built as
\begin{equation*}
\underbrace{\square \, \square \, \square}_{\text{serial ID}} \textendash \underbrace{\square \, \square \, \square \, \square \, \square}_{\text{product ID}}
\end{equation*}
where~$\square$ stands as a placeholder for an alpha-numeric character. Each serial ID may have different product IDs that fall under it to denote specific versions of a product or even closely related products. For example, the item in the first row of~Fig.\ref{fig:features} would have a serial ID of \texttt{LF1} and a product ID of \texttt{00018} and this particular serial ID is also associated with distinct~69 [00001 to 00069] product IDs (not shown in the Figure).

To start with, we consider all part numbers in the catalog and store them in a key-value structure where keys correspond to serial IDs and values correspond to the list of product IDs that fall under this serial ID.  This step is done offline and results in a hash map with about~$\approx 5899$ serial IDs each with an average of~$\approx 14$ product IDs.

At run-time, given a user query, the first step consists of assessing if it matches the pattern of characters that define a part number.  If it does, we split the query on the separator character and look up its serial ID in the offline dictionary.  Assuming that this key does exist, we then return a list of~10 items ranked according to their longest common subsequence with the user's query.  This ensures that if there is an exact match between the query and an existing item, this item will be returned first but we also show a few other related products as those may also be of interest to the seller.  

This section described the ``part number pattern matching'' module shown in the overall architecture in~Fig.\ref{fig:architecture}.  If any of these dictionary look-up or part number matching fails, the search system continues forward with the ``dynamic suggestions'' module which is the topic of the next section.

\section{Dynamic suggestions}
\label{sec:Suggestions}

\begin{figure}[ht]
\begin{center}
\centerline{\includegraphics[width=0.7\columnwidth]{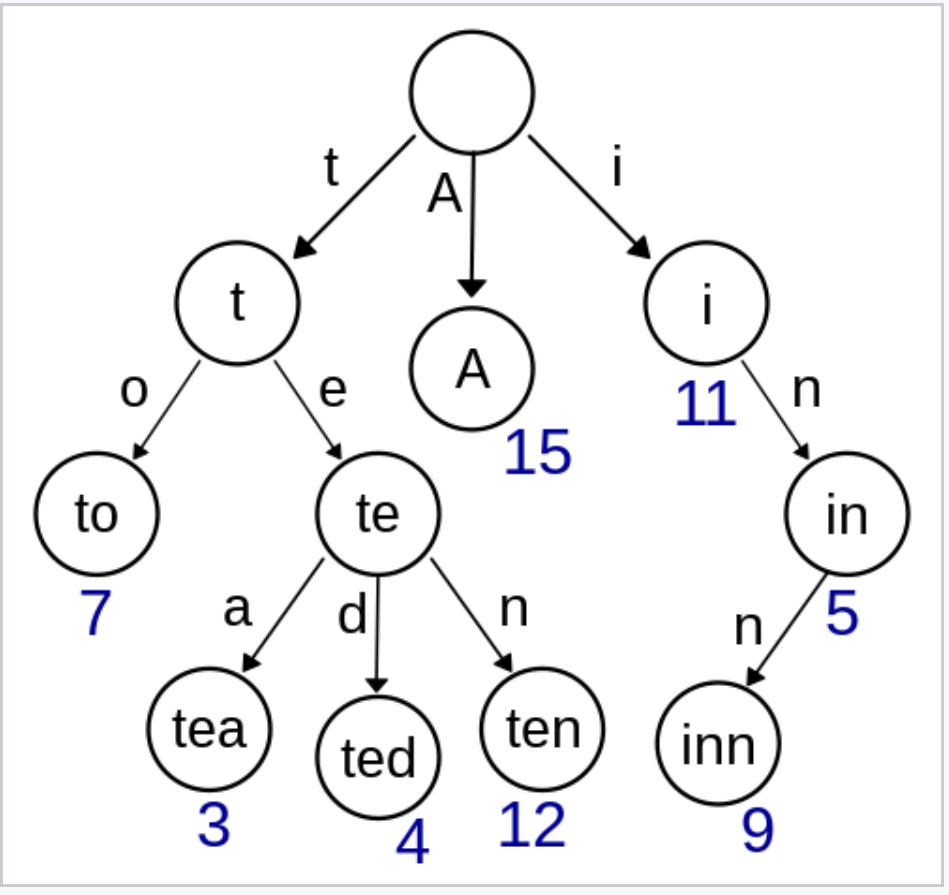}}
\caption{Example of a Trie data structure storing 10 strings~\cite{Trie}.  Notice how unlike a binary search tree, all the children of a node have a common prefix of the string associated with that parent node.  As soon as the user types (for example), the letter ``t'', the Trie can immediately suggest~4 possible matches corresponding to leaf nodes~7,~3,~4 and~12.  As the user continues to type more characters the space of possible matches rapidly (exponentially) decreases.}
\label{fig:Trie}
\end{center}
%\vskip -0.4in
\end{figure}

It is intended that search results should be presented in the form of dynamic suggestions as soon as users begin typing characters into the search box.  Furthermore, those results should be refreshed immediately after each new character.

Trie data structures~\cite{Trie1, Trie2} are a well-suited solution for handling dynamic search due to their retrieval complexity growing linearly with the number of characters in the query, independent of the number of nodes in the Trie.  As users type more characters, the Trie efficiently navigates through the branches, instantly offering relevant suggestions even for very large datasets.  An example of a simple Trie data structure is presented in~Fig~\ref{fig:Trie}.

In our case, we have populated the leaf nodes of the Trie with the part numbers, item names and friendly names (when available) resulting in~$\approx 228,000$ leaf nodes.  Overall the Trie occupies~$\approx 50 \, \text{MB}$ of memory.  Its performance is discussed in Section~\ref{sec:performance} and its place in the overall search architecture is explained in~Fig.\ref{fig:architecture}.

Successful Trie retrieval assumes that the order in which characters are typed form an exact subsequence match to at least one of the strings stored in the leaf nodes of the Trie.  Although there have been attempts to generalize Tries~\cite{bergman, Wenbo}, we have limited the scope of our current implementation to exact subsequence matches so that the size of the resulting Trie remains rather small and reserve fuzzy subsequce matches as a future extension.

\section{User-activated search}
\label{sec:completeSearch}

\begin{figure*}[ht]
\begin{center}
\centerline{\includegraphics[width=1.94\columnwidth]{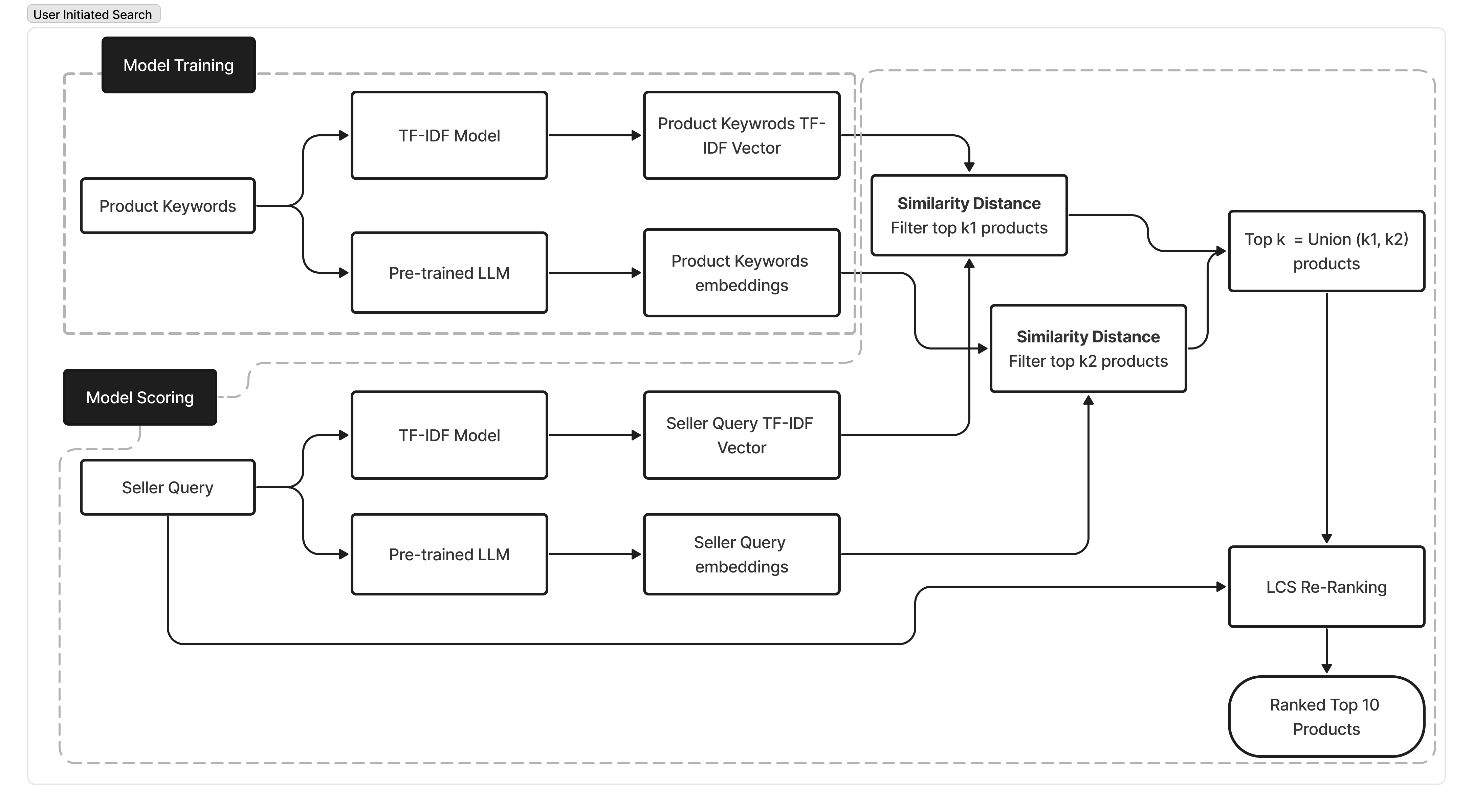}}
\caption{User-initiated search. {\bf Model training)} We start by encoding the catalog of products using both a pre-trained LLM and into character-level TF-IDF vectors. In our case, the pre-trained~LLM was chosen to be~\texttt{multi-qa-MiniLM-L6-cos-v1}~\cite{miniLM} as explained in~Section~\ref{sec:completeSearch} but others may be used as drop-in replacements. Those representations of the items in the catalog are then cached into memory.  {\bf Model scoring)}  Once the user pushed the ``search'' button, the seller query is transformed into a character-level TF-IDF vector as well as a second vector obtained by running this query into a pre-trained LLM.  At this point, the process resembles the model training phase with the only difference that we are now vectorizing only a single seller query at the run-time of the search.  Once both embedding vectors are obtained, we compare them respectively with the product embeddings we had stored during the model training phase.  We decide to keep the top~$k_1$ (resp. top~$k_2$) highest cosine similarity pairs between the character-level TF-DF embeddings (resp.~LLM embeddings). Given that there may be duplicates between both lists, we keep only the top-$k$ (where $k \leq k_1 + k_2$) of their union.  This final list of~$k$ product candidates is finally re-ranked according to the highest LCS scores between the candidates and the original seller query.}
\label{fig:userSearch}
\end{center}
%\vskip -0.4in
\end{figure*}

While the previous section shows how  dynamic suggestions can be returned quickly to the user, those results do not consider contextually similar terms.  

\subsection{Spell checker}
\label{sec:spellChecker}

Once the query is completely written into the search box, we start by processing it using the~\texttt{pyspellchecker}~\cite{spellCheck, Norvig} library.  This sophisticated spell checker works by comparing permutations within a predefined Levenshtein distance based on a set of default dictionaries.  

As can be seen in the~3$^\text{rd}$ row of Table~\ref{tab:perf}, the query pre-processing with the spell checker significantly improves the performance by almost~$4\%$.  In our case, we have noticed an even slightly better performance by enhancing the default dictionaries of~\texttt{pyspellchecker} with the~$\approx 540$ known abbreviations discussed in~Section~\ref{sec:abbreviationExpansion}.

\subsection{Similarity between vector representations of queries and items}
\label{sec:similarity}

In order to extend the scope of possible results returned by our search system, we resort to two types of information retrieval techniques:

\begin{itemize}
    \item The first one is based on the classical Term Frequency - Inverse Document Frequency (TF-IDF)~\cite{tfidf} albeit implemented at the character-level.  The output of the TF-IDF on a list of characters is a vector~\footnote{Overloading common ML terminology, we will refer to this vector as an embedding since it serves the same purpose as LLM embeddings.} of dimensionality~5000 (see Table~\ref{tab:perfTFIDF} for performance comparison for other embedding dimensions).  Ranking the cosine similarity between the TF-IDF embedding and the product embeddings generates a list of~$k_1$ product candidates that best resemble the user query. (In practice we have used~$k_1 = 50$.)

\begin{table}[tp]
%\vskip 0.05in
\begin{center}
\begin{small}
\begin{sc}
\begin{tabular}{|l|l|l|}
\toprule
Search method & Accuracy & Latency (ms) \\
\midrule \midrule
Word TF-IDF ($\sim 2000$) & 47.86 & 66.2 \\
Word TF-IDF ($\sim 5000$) & 53.69 & 153.2 \\
Word TF-IDF ($\sim 10000$) & 49.61 & 277.7 \\
\midrule
Char TF-IDF ($\sim 2000$) & 72.76 & 63.4 \\
Char TF-IDF ($\sim 5000$) & 75.09 & 153.1 \\
Char TF-IDF ($\sim 10000$) & 68.09 & 268.8 \\
\bottomrule
\end{tabular}
\end{sc}
\end{small}
\end{center}
\caption{Performance for different configurations of TF-IDF. We experimented with embedding dimensionalities~$\{ 2000, 5000, 10000 \}$.  One can see that the peak performance is obtained with~5000.  We note also that the latency starts to approach the (real-time) requirement of~$\sim 300$ms we need to satisfy.  We also experimented with different granularities, word vs. character, of the TF-IDF.  As explained in the main text, character-level TF-IDF is far more accurate than its word-level counterpart. Latency is reported in milliseconds (ms).}
%\vskip 0.1in
\label{tab:perfTFIDF}
\end{table}

    As discussed in more detail in Section~\ref{sec:performance}, because of the abbreviations and relatively short queries in CRMs, it is important to create character-level representations rather than word level as is more common in the general-purpose search literature.
    
    \item The second one is based vector similarity of pre-trained Large Language Model (LLM) embeddings. In this work we use the symmetric~\texttt{multi-qa-MiniLM-L6-cos-v1}~\cite{miniLM} model which maps input text into~384-dimensional embeddings.  This model is trained on~$\approx 215,000$ question/answer pairs from various sources and domains, including StackExchange, Yahoo Answers, Google \& Bing search queries and many more.  Similarly to the first technique, we rank the cosine similarity between the user query and the product embeddings to generate another list of~$k_2$ product candidates that best resemble the user query. (In practice we have used~$k_2 = 50$.)

\begin{table}[t]
%\vskip 0.05in
\begin{center}
\begin{small}
\begin{sc}
\begin{tabular}{p{0.4\linewidth}|p{0.18\linewidth}|p{0.22\linewidth}}
\toprule
Search method & Accuracy & Latency (ms) \\
\midrule \midrule
multi-qa-MiniLM-L6-cos-v1 & 35.99 & 48.1\\
\hline
GloVe 840B & 33.65 & 6.34  \\
\hline
word2vec-GoogleNews & 28.21 & 5.41 \\
\hline
msmarco-distilbert-base-v4 & 24.31 & 85.29\\
\bottomrule
\end{tabular}
\end{sc}
\end{small}
\end{center}
\caption{We experimented with different LLMs from the sentence transformers HuggingFace library. It turns out that~\texttt{multi-qa-MiniLM-L6-cos-v1}~\cite{miniLM} which is a symmetric encoder gives the best performance compared to~\texttt{msmarco-distilbert-base-v4}~\cite{MSMarco} which is an asymmetric encoder.  We also report the performance of simpler embedding techniques such as~\texttt{GloVe 840B}~\cite{Glove1, Glove2} and~\texttt{word2vec-GoogleNews}~\cite{word2vec}.  Because of their relative simplicity, those embeddings are able to run in much faster latencies but still display a significant performance gap with respect to~\texttt{multi-qa-MiniLM-L6-cos-v1}. Latency is reported in milliseconds (ms).}
%\vskip 0.1in
\label{tab:perfLLMs}
\end{table}
    
Although LLM embeddings have grown immensely in popularity recently, the performance of this technique lags behind TF-IDF, in our case, for reasons explained in Section~\ref{sec:performance}.  We use them here as a supplement that brings in additional performance for longer, more structured queries that are not very frequent in CRM product search queries. A few different experiments are presented in Table~\ref{tab:perfLLMs} where we also report the performance of more traditional embeddings such as~\texttt{GloVe 840B}~\cite{Glove1, Glove2} and~\texttt{word2vec-GoogleNews}~\cite{word2vec}.
\end{itemize}

A complete illustration of how TF-IDF and LLM embeddings are connected together is offered in~Fig.\ref{fig:userSearch}.  Contrary to the dynamic suggestions of Section~\ref{sec:Suggestions}, both of these techniques require to have the complete user query typed into the search box.  The search process itself is initiated when the user clicks on a ``search button''.

\subsection{Ensemble re-ranking}

Taking the set union of the two lists of~$k_1$ (resp.~$k_2$) product candidates generated via TF-IDF (resp.~LLM) embeddings produces a final list of~$k$ candidates.  Note that~$k < k_1 + k_2$ since there may be duplicated candidates.  In practice we have found that this yields an average of~$k \approx 90$ candidates.

The final step consists in re-ranking the~$k$ candidates.  We do this by sorting according to the pairwise normalized longest common subsequence (nLCS) score between the original user query and the actual product names (i.e. not embeddings) of the~$k$ candidates.  LCS normalization is done by dividing the length of each LCS with the length of the user query.  This way a nLCS of 1 indicates a perfect match not only in LCS but also in the length of the candidate and the query.  This step is illustrated as the ``LCS re-ranking'' box in the bottom right of~Fig.~\ref{fig:userSearch}.

\subsection{Complete search system architecture}
\label{sec:Architecure}

Combining together the part number pattern matcher (see~Section~\ref{sec:patternMatch}) with the dynamic suggestions (see~Section~\ref{sec:Suggestions}) and the user-activated search (see~Section~\ref{sec:completeSearch}) under a single architecture presents a powerful solution for handling a more thorough set of search scenarios that happen in CRM product searches.

\begin{figure*}[ht]
\begin{center}
\centerline{\includegraphics[width=2\columnwidth]{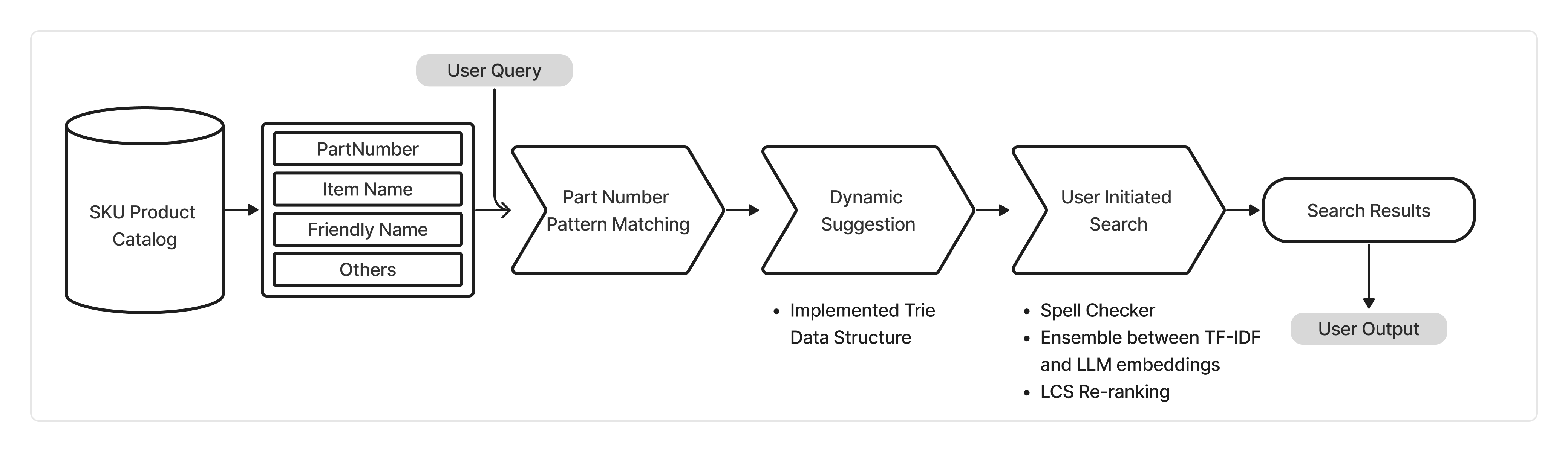}}
\caption{High-level summary of the complete SKU product search architecture for CRMs.  The first component consists of the part number pattern matcher discussed in~Section~\ref{sec:patternMatch}.  In case, the search query cannot be handled by this module, it moves on to the dynamic suggestions detailed in~Section~\ref{sec:Suggestions}.  Finally, the last component consists of the user-initiated search which is detailed in Section~\ref{sec:completeSearch} and Fig.~\ref{fig:userSearch}.}
\label{fig:architecture}
\end{center}
%\vskip -0.4in
\end{figure*}

Since all of the other constitutive components have been covered in the previous relevant sections, we now refer the reader to~Fig.~\ref{fig:architecture} where the overall architecture connecting all these individual components is presented.

The next section will be dedicated to the performance evaluation and ablation study of the search system presented in~Fig.~\ref{fig:architecture}.

\section{Fast and slow, the accuracy / latency trade-off}
\label{sec:performance}

\begin{figure}[ht]
\begin{center}
\centerline{\includegraphics[width=\columnwidth]{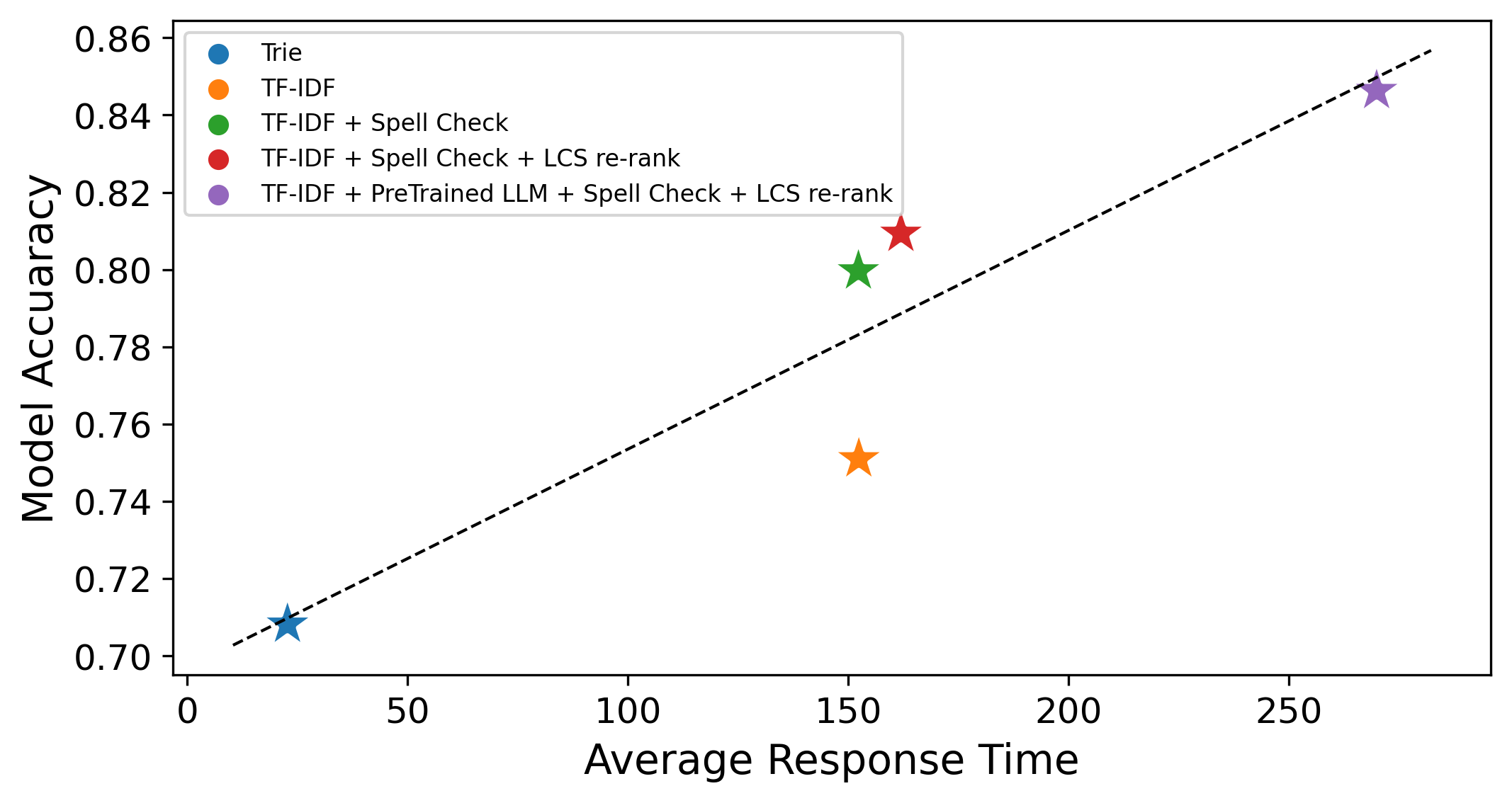}}
\caption{Illustration of the trade-off between response time and accuracy of the model. The dashed black line illustrates the linear growth in model accuracy as the response time gets longer when more modules are added into the overall architecture.}
\label{fig:modelPerformance}
\end{center}
%\vskip -0.4in
\end{figure}

\begin{table*}[h]
%\vskip 0.05in
\begin{center}
\begin{small}
\begin{sc}
\begin{tabular}{|l|l|l|l|}
\toprule
& Search method & Accuracy & Latency (ms) \\
\midrule \midrule
1 & Trie & 70.82 & 22.7  \\
2 & TF-IDF & 75.01 & 152.38  \\
3 & TF-IDF + Spell Check & 79.96 & 152.29  \\
4 & TF-IDF + Spell Check + LCS re-rank & 80.93 & 161.93  \\
5 & TF-IDF + PreTrained LLM + Spell Check + LCS re-rank & 84.63 & 269.88  \\
\bottomrule
\end{tabular}
\end{sc}
\end{small}
\end{center}
\caption{Summary of the selected search modules from~Section~\ref{sec:Suggestions} and Section~\ref{sec:completeSearch} that we used in the overall search architecture.  The same data is also represented visually in Fig.~\ref{fig:modelPerformance}. Latency is reported in milliseconds (ms).}
%\vskip 0.1in
\label{tab:perf}
\end{table*}

In order to evaluate the performance of our architecture, we manually labeled 513 sellers' queries that in the existing MSX search did not return anything.  However, upon closer examination with domain experts, we discovered that those queries should have returned a relevant SKU existing in the catalog.  So we manually pair those queries with the correct SKU that the seller intended to see and then we can evaluate the performance of the search system on those queries.

Accuracy is measured as the average (over those 513 manually labeled search queries) number of times the correct SKU is returned as part of the top-10 search results.  Mean Reciprocal Rank – MRR is also a popular evaluation metric.  In practice we have observed a very high degree of correlation between MRR and the accuracy so, for the sake of simplicity, the performance tables report only the accuracy (success@10) as defined above.
 
With our search solution deployed as an~Azure Machine Learning - AML endpoint, latency is quantified simply as the average response time of the~API.  Requirements from our stakeholder teams dictate that the search latency should not exceed~300ms.

Some noteworthy observations from Table~\ref{tab:perf} and~Fig~\ref{fig:modelPerformance} are that:
\begin{itemize}
    \item As expected, the rapid nature of dynamic search is associated with lower accuracy whereas complete search ensures a more thorough and accurate retrieval of information by allowing the search engine more time to process the query.
    \item As already discussed in Section~\ref{sec:similarity} and Table~\ref{tab:perfTFIDF}, the performance of the word-level TF-IDF is very poor with an accuracy of only about~$\approx 54\%$.  This is due to the nature of CRM product SKUs which tend not to be words and also typically contain many abbreviations.
    \item It is interesting to note that removing the character-level TF-IDF and relying entirely on the LLM embeddings results in a very poor performance.  This is due to the fact that LLMs are typically trained on longer text that resemble more English queries than the short abbreviated product names sellers tend to search for in CRMs. The fusion between character-level TF-IDF and LLM embeddings allows for a more nuanced understanding of user intent and enables the system to retrieve results that align better with the user's query. Note that those long queries represent less than~2\% of the real-world search queries used by sellers so that the bulk of the performance is given by the~TF-IDF branch and not the~LLM one; We discuss in the conclusion future~LLM extensions that will attempt to correct this.
\end{itemize}

\section{SKU description generation}
\label{sec:GPTgen}

Although it is not directly part of the search algorithm itself it is tremendously helpful for sellers to see a full-fledged English description of the SKUs when they review the search results.

We performed iterative prompt engineering based on the chat completion API of GPT-3.5-Turbo.   In the absence of available labeled data we used zero shot learning for generating descriptions from the LLM.  We designed prompts in a way that the model can generate a summary for a SKU without any specific examples or training data related to that particular SKU.  We set the maximum token limit for the response to control the length of the output and reduce variability.   This allows us to generate deterministic SKU descriptions for the entire MSX catalog.  Using low-temperature-settings to further improve responses, we make another API request to summarize the description in a maximum of~250 characters.  

We observed an average latency of~2.7s (max of 7s) per API request.  With such a high API response time, processing~$87,000$ MSX catalog items is network bound and may take considerable time, 60+ hours, if implemented na\"ively.  We optimized using parallel processing of sending 20+ concurrent API requests to complete the entire process in less than~120 minutes (excluding IO time of writing to ADLS).

As an example, for part number LF1-00018 which we show in Fig.1, the SKU description comes out to be: \textit{``The Surface Laptop 4 is a 13-inch device with an i7 processor, 16GB RAM, and 512GB storage. It comes with Windows 11 and is designed for commercial use. The hardware is in Ice Blue color and the SKU is AUSTL/NZ''}.  We can clearly see the description is much more easily readable for the seller than the Product Friendly Name and Item Name from Fig.1. for the same part number and helps the seller in picking the right SKU/part number quickly.

\section{Conclusion}

In this paper, we show that by combining traditional technique of character-level TF-IDF with pre-trained LLMs, we can significantly reduce the rate of ``zero search results'' and ``too many results'' problems which exist for SKU search in today's CRM systems.  This architecture is now pending production deployment in the Dynamics CRM where it is expected to improve seller satisfaction ratings and save time for sellers when searching for products.

%Once we deploy this solution to production and measure the actual results, we intend to share the results on improvement in Click-Through rate (CTR), reduction in ''Zero Search Results' and 'Too many results' and improvement in Seller Satisfaction rating and time saved per opportunity by the seller in searching for the relevant products.

By offering a system that relies on multiple and orthogonal search modules, our search solution caters to varying user preferences and needs.  Based on the search latency results presented here, we expect our top level accuracy model to perform at least as fast as the existing CRM search once the model is integrated to the enhanced overall CRM system.

Although one would also like to consider more modern and potentially more accurate embeddings such as those offered by OpenAI~\cite{openAI}, we have to satisfy strict latency requirements of providing search results within~300~ms for a peak volume of approximately~$\approx 20,000$ daily searches.  At this point, those Service Level Agreements – SLAs imposed on us by our stakeholder teams, coupled with our own Azure subscription quotas, exclude us from leveraging OpenAI embeddings though we hope to explore this path more going forward.

In future work, we intend to investigate possible lift in performance by fine-tuning the language models instead of relying on their pre-trained versions.  In  addition, we intend to develop our own semantic abbreviation embeddings specifically trained on abbreviations so we can capture some semantic similarity, not at the word / sentence level but rather at the abbreviation level.  We also plan to explore ways to automatically discover abbreviations and their expansion rules~\cite{autoAbbreviatons1, autoAbbreviatons2}.

\section{Acknowledgements}

We thank our colleagues in CX Data and SPS for their feedback and support. In particular, we thank the ML Review board constituting of Daniel Yehdego, Vijay Agneeswaran and Ivan Barrientos for their valuable inputs on this model. We also thank Manpreet Singh for his inputs on pre-trained LLM models and in deploying the real-time AML endpoint. We thank the SPS PM team of Ullas Gupta and Melinda Amoratis for bringing this key MSX Search problem to us and continuously evaluating our solution and getting us feedback from domain experts. We also thank Chandra Mouleswar, Dhana Vujjini and Lony Choudhury from SPS engineering team in deploying this solution in MSX.

%\newpage
\bibliographystyle{ieeetr}
\bibliography{MainText}

\end{document}